\documentstyle[epsfig]{aipproc}

\begin{document}
\begin{flushright}
UR-1517\\
ER/40685/912\\
hep-ph/9802215\\
January 1998\\
\end{flushright}

\title{Gluon Radiation in Top Production and Decay at Lepton 
Colliders\footnote{Presented at the Workshop on Physics at the 
First Muon Collider and at the Front End of a Muon Collider, Batavia, IL, 
Nov.~6--9, 1997; work supported in part by the U.S. Department of Energy.}}

\author{Lynne H. Orr$^*$}
\address{$^*$Department of Physics and Astronomy\\
University of Rochester\\
Rochester, NY 14627-0171}

\maketitle

\begin{abstract}
In this talk we discuss gluon radiation in top production and decay.
After reviewing results for hadron colliders we consider soft gluon
radiation at lepton colliders and present gluon distributions that
are potentially sensitive to production-decay interference effects.
\end{abstract}

\section*{Introduction}

Top quark events are often accompanied by gluons.  This gluon radiation
can occur in association with both the top production and top 
decay processes, and must be understood if we are to make sense of 
top physics.  This is especially true for top momentum reconstruction, where
a jet originating as a gluon may or may not be a top quark decay product.
Top momentum reconstruction can play a crucial role not only in precision top
mass measurements but also in identification of top events by using a mass 
cut.  In addition, the pattern of the gluon radiation itself contains 
information about the top production and decay processes and their
color structure, and can be sensitive to new physics. 

In this talk we discuss results for gluon radiation in top events at hadron
colliders, and then focus on lepton colliders, noting some similarities and 
differences along the way.  Further details can be found in the references.

\subsection*{Review of Hadron Collider Results}

In top production and decay at hadron colliders, strongly interacting
particles in the intial, intermediate, and final states all give rise to 
gluon radiation.  In reconstructing the top mass, it matters whether
jets from these gluons are part of the top production or decay process.
If the gluons arise in top production, they should be ignored in 
mass reconstruction, but if they are part of the top decay process
then they must be included.\footnote{Interference between gluons 
from production and those from decay is negligible provided their
energies are large compared to the top width, a reasonable assumption 
for gluons that are to be detected as jets at hadron colliders.  See
the lepton collider section for further discussion.}  

In practice that seemingly straightforward rule is hard to follow for 
several reasons.  First, gluon
jets are not necessarily distinguishable from other jets from top 
decays.  Second, even if the gluon jets can be identified one 
cannot easily distinguish between production- and decay-stage radiation.
In a study of gluon radiation in top production and decay at the 
Tevatron \cite{osstev} we found that for kinematic cuts meant 
to mimic typical detector capabilities, the amounts of production-
and decay-stage radiation were comparable (ignoring radiation from
$W$ decay products).  Furthermore, the two 
were not easily separated.  Production-stage radiation is well
spread out in rapidity, and although decay-stage radiation is more 
central, there is significant overlap in the relevant detector regions.
Even proximity to one of the $b$-quark jets is not sufficient to 
uniquely identify a gluon as being associated with decay, for example.
This underscores the importance of understanding gluon radiation in
top processes.

\begin{figure}[b!] 
\centerline{\epsfig{file=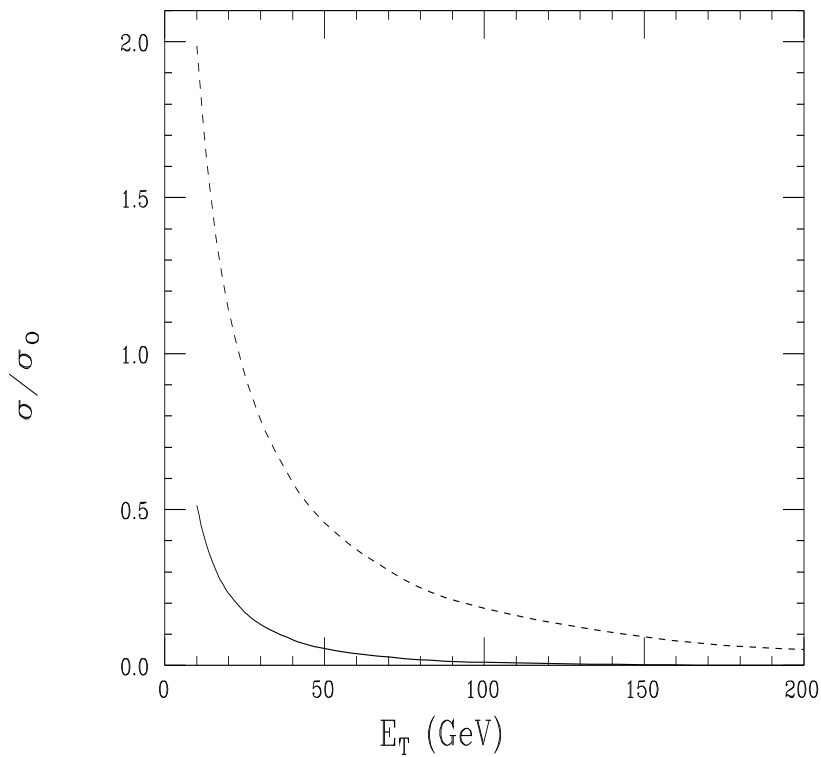,height=5.in,width=5.in}}
\vspace{10pt}
\caption[f1]
{Ratio of $\sigma(t\bar t j$) to $\sigma(t \bar t)$ for $E_T^j >E_T$
at the Tevatron (solid line) and LHC (dashed line).\cite{osslhc}} 
\label{orrfig1}
\end{figure}

At the LHC $pp$ collider, the situation is slightly different.\cite{osslhc}
Its higher energy and luminosity give a vast increase in the top
production cross section, but at the same time the amount of gluon
radiation is also increased, notably at the production stage. 
This is illustrated in Figure \ref{orrfig1}, which shows the 
ratio of cross sections for $t \bar t j$ and 
$t \bar t$ production at the Tevatron (solid line) and LHC
(dashed line) as a function of the minimum transverse energy 
$E_T$ of the gluon.  The LHC is seen to have a vast increase in 
production-stage radiation over the Tevatron for any given 
$E_T$ cut.  

In contrast, the decay-stage radiation at the LHC looks similar to that at 
the Tevatron, because the kinematics of the produced top quarks are 
similar.  The main difference is that the $t$'s are more spread out
in rapidity at the LHC, which gives rise to a decay-stage gluon distribution
that is also more spread out.  

At the LHC, then, the individual distributions of production- and 
decay-stage radiation are similar to those at the Tevatron, but 
production-stage radiation dominates by far.  This can be a problem,
for example, when those gluons overlap with jets from $t$ decays,
causing mismeasurement of parton energies that can feed into
momentum reconstruction.  Such effects cannot be avoided but they can be
taken into account as long as they are incorporated into the relevant analyses.

\section*{Gluon Radiation at Lepton Colliders}

At lepton colliders there is no gluon emission from the initial state. 
That means that, for purposes of studying gluon radiation in top production
and decay, there is
no difference between $e^+e^-$ and $\mu^+\mu^-$ colliders.  (There are 
important differences in, for example, precision studies at 
the $t \bar{t}$ threshold; see the talk by M.~Berger.\cite{bergertalk})
The absence of initial-state gluon radiation and the relatively
simpler kinematics compared to hadron colliders allows for the 
possibility of more detailed study of radiation patterns at lepton
colliders.  The pattern of gluon radiation can give information about the 
dynamics of the underlying process, including interplay between top 
production and decay, and possible signals of new physics.

\subsection*{Soft Gluon Radiation}

The study of soft gluon radiation patterns in high energy process, 
sometimes called 
``partonometry,'' is useful for a number of reasons.  The infrared
singularity in the cross section means that there is a high probability
for emitting soft gluons, whose distribution determines the distributions of 
soft hadrons or minijets in such events.  Studying these distributions
then gives information about the underlying process.  In doing calculations,
the soft limit is useful because the cross section factorizes into 
the lowest order (differential) cross section and a piece due solely
to the gluon radiation: 
\begin{equation}
dN \equiv 1/\sigma_0 d\sigma_g =
 \frac{dE_g}{E_g}\frac{d\Omega}{4\pi}\>\frac{C_F\alpha_s}{\pi}\>\cal{R}\>,
\end{equation}
where $E_g$ and $\Omega$ denote the gluon energy and solid angle.

In $t \bar{t}$ production and decay at lepton colliders, although there is
no initial state gluon radiation, there are still gluons emitted in both the 
top production and decay stages (respectively before and after the top quark 
goes on shell).  It is straightforward to find $\cal{R}$
and decompose it into the contributions corresponding to radiation from
the various stages; see \cite{kosnpb}.  

\begin{figure}[b!] 
\centerline{\epsfig{file=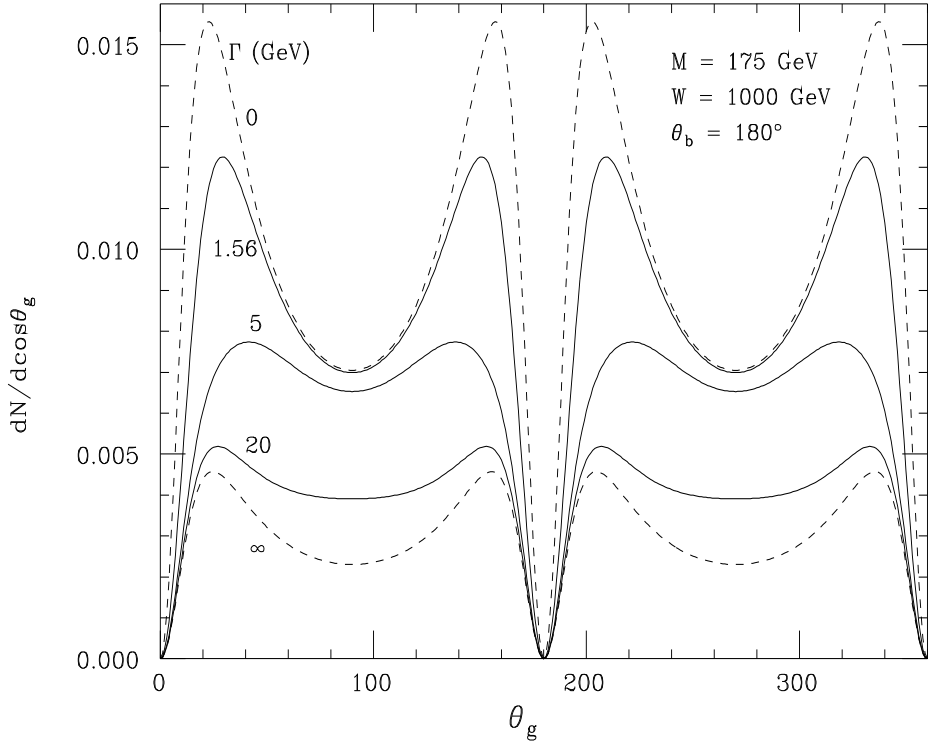,height=5.in,width=5.in}}
\vspace{10pt}
\caption[f1]
{Soft gluon distribution in top production and decay at lepton colliders 
as described in the text, for $t$'s decaying to backward $b$'s and 
collision energy $1\ {\rm TeV}$.} 
\label{orrfig2}
\end{figure}

One interesting result of such
an analysis is that there can be  interference between production- and
decay-stage gluons when the gluon energy is roughly comparable to the 
top width $\Gamma$.  This can be understood heuristically in terms of
Breit-Wigner distributions for the square of the top momentum with and 
without the gluon, corresponding to radiation in the decay and 
production stages, respectively.  The peak separation
is roughly the gluon energy $E_g$, and since each distribution has width 
$\Gamma$, when $E_g \sim \Gamma$, the two distributions overlap, giving 
rise to non-negligible interference.\footnote{We can now see why the 
production-decay interference is negligible at hadron colliders.  We
consider extra jets with minimum transverse energies of $20$ and 
$40\ {\rm GeV}$
at the Tevatron and LHC, respectively; such energies are much larger
than the Standard Model top width of about $1.5\ {\rm GeV}$.}
Because the presence of this interference for a given gluon energy 
depends on the value of $\Gamma$, it follows that gluon
distributions can be sensitive to the top width.

\begin{figure}[b!] 
\centerline{\epsfig{file=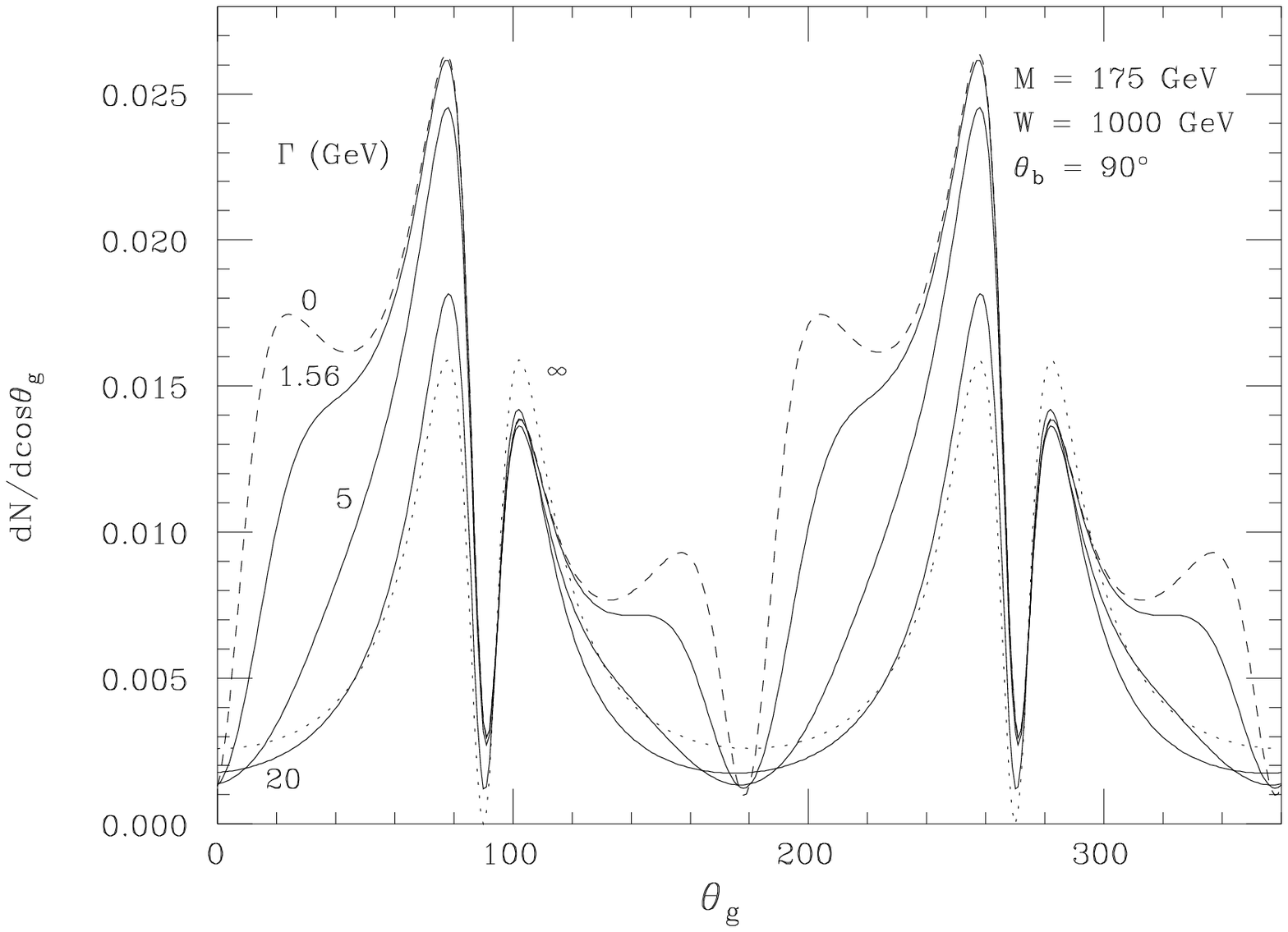,height=5.in,width=5.in}}
\vspace{10pt}
\caption[f1]
{Soft gluon distribution in top production and decay at lepton colliders 
as described in the text, for $t$'s decaying to $b$'s  at $90^\circ$ and 
collision energy $1\ {\rm TeV}$.} 
\label{orrfig3}
\end{figure}

In Ref.~\cite{kosnpb} the gluon distribution $\cal{R}$ was studied 
for specific kinematic configurations for a top mass of $140\ {\rm GeV}$.
Here we update those results for a $175\ {\rm GeV}$ top and for
energies appropriate to muon collider studies.  In Figure \ref{orrfig2}
we show the gluon distribution for center of mass energy $1\ {\rm TeV}$
for a configuration where the $t$ quark decays to a backwards $b$ and 
the $\bar t$ also decays to a backwards $\bar b$.  The distribution is
shown as a function of $\theta_g$, the angle between the gluon and 
the top quark.  For this configuration, the $t$ and $\bar b$ quarks
are located at $\theta_g=0^\circ$ and the $\bar t$ and $b$ quarks are
at $\theta_g=180^\circ$.  The gluon energy is taken to be $5\ {\rm GeV}$.
The overall shape shows the ``dead cone'' behavior characteristic 
of radiation from heavy quarks --- radiation along the quark 
direction is suppressed by the mass.
The curves are for different values of the top width $\Gamma$, as 
labeled.   
We see that radiation in the SM case ($\Gamma=1.5\ {\rm GeV}$) 
is suppressed compared to that for $\Gamma=0$, due to destructive interference
between the production and decay stages.  As the width increases so does this 
suppression, and in the limit $\Gamma \rightarrow \infty$, the 
gluon distribution is what it would be if the $b$ quarks were 
produced directly:  the contribution of radiation from the top 
quark disappears altogether.

Top quarks at high energies are not terribly likely to decay to backward 
$b$'s, but interference effects remain if we look at a more likely 
configuration in Figure \ref{orrfig3}, where the $b$ quarks come out at 
$90^\circ$ to their parent top quarks.  Here the $t$ and $\bar t$ quarks
are at $0^\circ$ and $180^\circ$, respectively, and the $b$ and $\bar b$ 
are at $90^\circ$ and $270^\circ$.  We see more interesting structure 
in this case, because the $b$ directions do not coincide with those of 
the $t$'s.  
Again we see destructive 
production-decay interference that increases with the top width,
and in the $\Gamma \rightarrow \infty$ limit (dotted curve) 
there is no evidence of the top direction in the distribution.

\begin{figure}[b!] 
\centerline{\epsfig{file=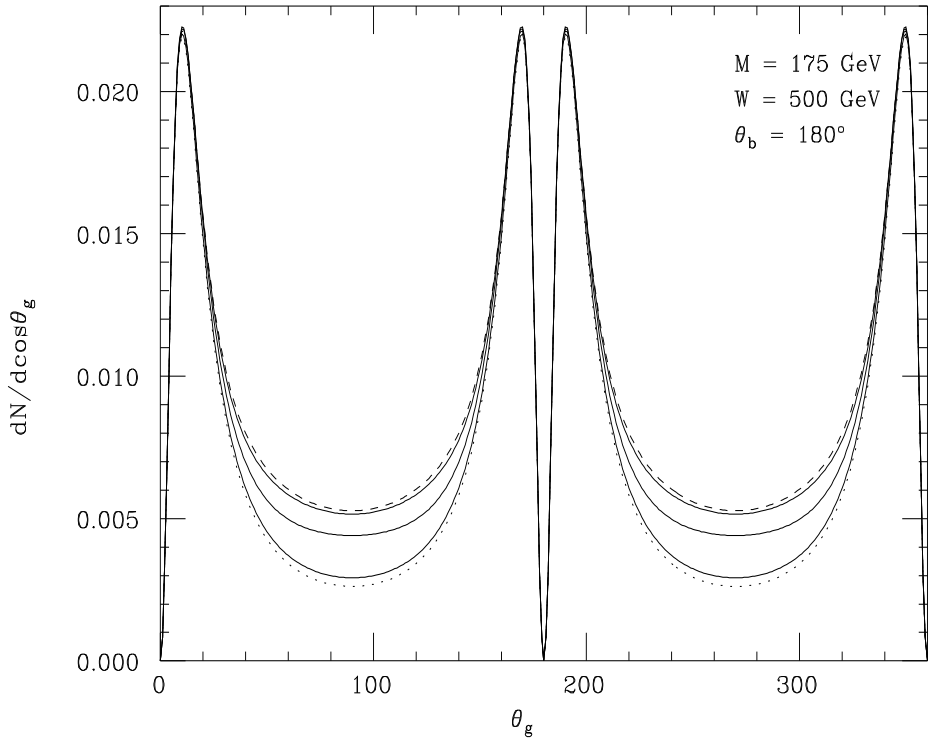,height=5.in,width=5.in}}
\vspace{10pt}
\caption[f1]
{Same as Figure \ref{orrfig2}, except for collision energy $500\ {\rm GeV}$.} 
\label{orrfig4}
\end{figure}

The $1\ {\rm TeV}$ collision energy discussed above is certainly desirable
from the point of view of a variety of physics topics.  However, as 
stated in the workshop guidelines, a first muon collider is more likely
to operate at the lower energy of $500\ {\rm GeV}$.  How do top 
production-decay
interference effects look in the gluon distribution at the lower
collision energy?
Not too promising, unfortunately, as can be seen in Figure \ref{orrfig4},
which is the same as Figure \ref{orrfig2} except that the collision energy
is $500\ {\rm GeV}$.  There is little sensitivity to the top width $\Gamma$.
The reason is that the top quarks come out with lower energy and 
the radiation is dominated by that from the $b$ quarks.  Results for the 
configuration corresponding to Figure \ref{orrfig3} show similar behavior.

At higher collision energies it appears that the distribution of soft
gluon radiation may be sensitive to the top quark width, which suggests
this as a method for measuring $\Gamma$.  This will be challenging, to 
say the least.  The results presented here are at the parton level only,
and moreover they are for fixed kinematic configurations.  A more 
realistic assessment requires a more detailed study of integrated 
cross sections.  The result is most likely to be that the sensitivity
leaves something to be desired.  However, it will still be
worth pursuing such measurements, because simply seeing the interference 
efects will be interesting.  Moreover, the width that appears here
is the {\it total} top width, independent of decay mode, and would serve
as a consistency check to compare to other measurements.  And it should 
be remembered that additional top decay modes, for example to supersymmetric
particles, are likely to increase the total width, which would make the effects
discussed here more easily observed.

\section*{Conclusion}

In summary, we have discussed issues associated with gluon radiation in top
production and decay at hadron and lepton colliders.   We noted that for
purposes of gluon radiation, there are no significant differences between
muon and electron colliders, since there is no initial state gluon radiation.
We presented some gluon distributions at lepton colliders that show
interference effects between production- and decay-stage gluon radiation,
and that are potentially sensistive to the value of the top quark width.
Finally, we note that the clean environment of lepton colliders allows for 
detailsed experimental studies of QCD effects in top quark physics, both 
within the Standard Model and beyond.

\end{document}